\documentclass[twocolumn,prb]{revtex4}
\usepackage{amsfonts}
\usepackage[T1]{fontenc}
\usepackage{amsmath,amsbsy,amssymb,graphicx}
\usepackage{times}

\begin{document}

\title{Topological switch 
\\
between second-order topological insulators and topological crystalline insulators}
\author{Motohiko Ezawa}
\affiliation{Department of Applied Physics, University of Tokyo, Hongo 7-3-1, 113-8656,
Japan}

\begin{abstract}
We investigate a topological switch between second-order topological insulators (SOTIs) and topological crystalline insulators (TCIs).
Both the SOTI and the TCI are protected by the mirror and inversion symmetries, for which we define the bulk topological numbers of the same type.
We take examples of square nanodisks on the square lattice and hexagonal nanodisks on the triangular lattice.
When inplane magnetic field is introduced parallel to one of the helical edges, the system becomes a  TCI.
The conductance along the edge is 1 in the unit of the conductance quantum $e^2/h$. 
As the inplane field is rotated, the conductance decreases as the gap of the edge states opens.
When it becomes orthogonal to a diagonal line, two topological corner states emerge on its vertices and the system becomes a SOTI.
When it becomes parallel to another edge, the system becomes again a TCI
and the conductance along the original edge becomes 0 but 1 along a new edge.
This may be used as a basis of a topological circuit changing switch.
Alternatively, the device may be used as a sensor to measure local magnetization on a sample surface with a resolution of 10 nm. 
\end{abstract}

\maketitle

\textit{Introduction:} Topological crystalline insulators (TCIs) are protected by crystal symmetries\cite{FuTCI,AndoReview}. A typical crystalline symmetry is the mirror symmetry.  In TCIs, topological boundary states emerge when the topology protecting crystal symmetry is preserved at the boundary. On the other hand, once the topology protecting crystal symmetry is broken at the boundary, the topological boundary states do not emerge. This is a generalization of the bulk-boundary correspondence protected by crystalline symmetry.

Recently, higher-order topological insulators attract much attention%
\cite{Fan,Science,APS,Peng,Lang,Song,Bena,Schin,FuRot,EzawaKagome,EzawaPhos,Gei,MagHOTI,Kha,HexaHOTI,Bis},
where a typical topology protecting symmetry is a rotational symmetry. 
For example, in a second-order topological insulator (SOTI) in two dimensions, 
the one-dimensional edge states are gaped but there emerge zero-dimensional corner states. 
This phenomenon occurs when the topology protecting symmetry is broken at the edges but preserved at the corners.

\begin{figure}[t]
\centerline{\includegraphics[width=0.45\textwidth]{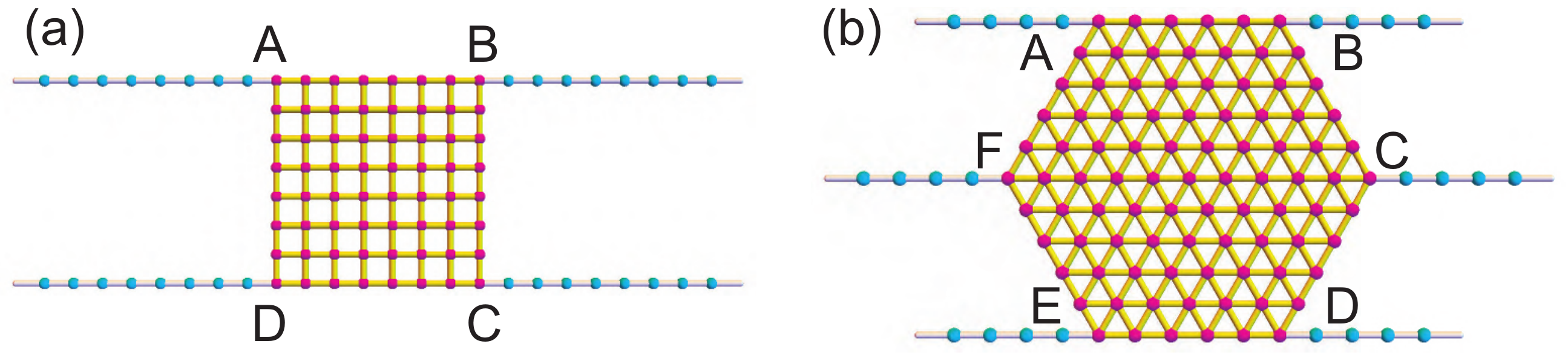}}
\caption{Illustration of topological circuit changing switch made of (a)
square and (b) hexagonal nanodisks. We attach four (six) semi-infinite leads
to the corners of the square (hexagonal) nanodisk. }
\label{FigIllust}
\end{figure}

There are apparently common features between the TCI and the SOTI.
In this paper, we investigate a topological switch between a TCI and a SOTI, by considering a system possessing the mirror and inversion symmetries.
We take examples of square and hexagonal nanodisks on the square and triangular lattices, respectively,
We have in mind a topological device realizing a circuit changing switch, as illustrated in Fig.\ref{FigIllust}, 
where the connectivity of the circuit is controlled by the direction of inplane magnetic field. 
The conductance is quantized and switched between 0 and 1 in the unit of the conductance quantum $e^2/h$. 
Indeed, when the field is parallel to one of the edges, there are topological edge channels carrying conductance quanta.
The gap of the helical edge opens when the field is not parallel to the edge.
When the field is orthogonal to a diagonal line, there are no topological edge channels but topological corner states emerge. 
The conductance has a sharp angle dependence on the direction of the field. 
It is shown that the system is a TCI when the field is parallel to the edge,
while the system is a SOTI when the field is orthogonal to a diagonal line.
The bulk topological number is defined and quantized in these cases.

\begin{figure*}[t]
\centerline{\includegraphics[width=0.95\textwidth]{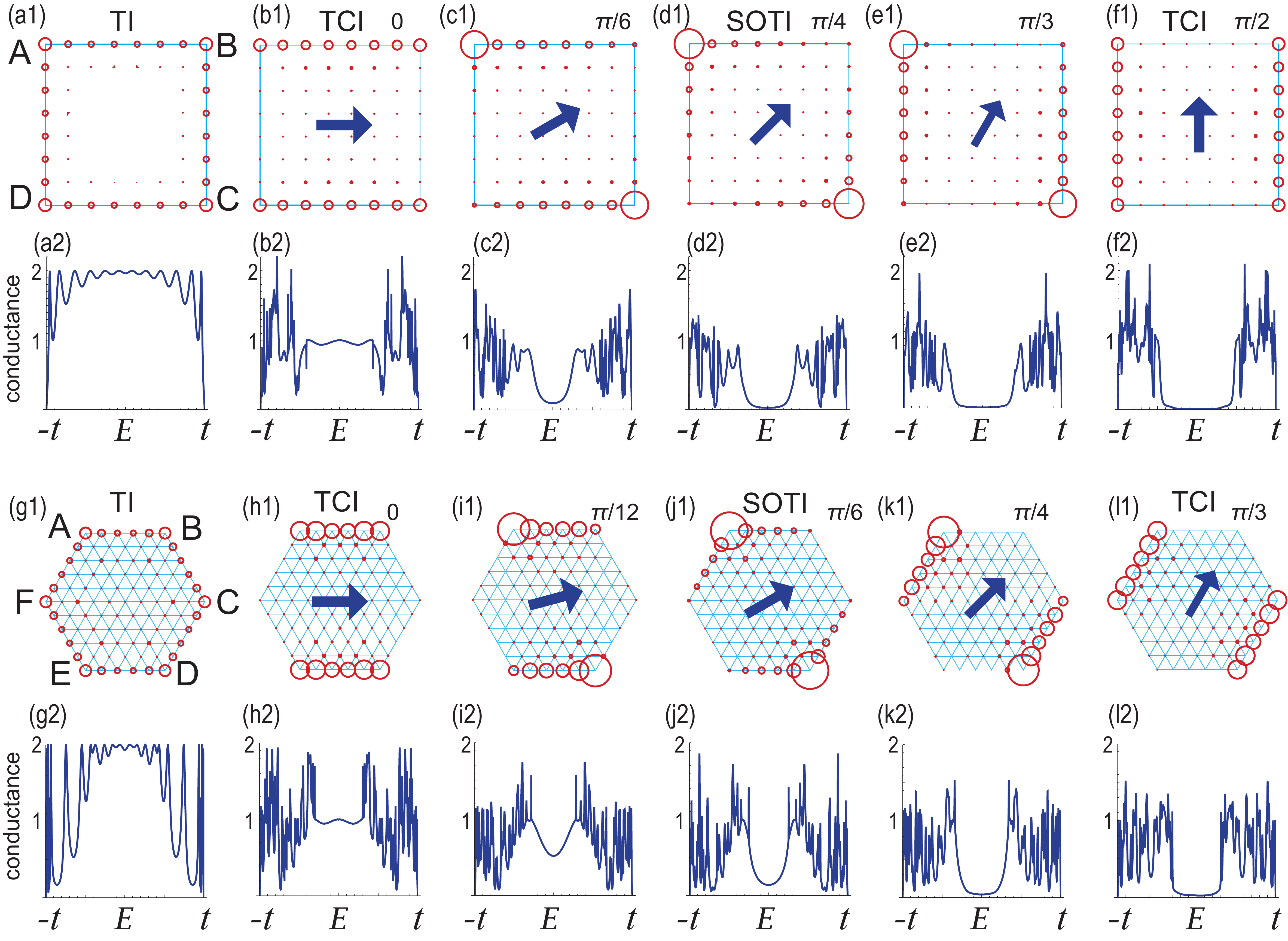}}
\caption{LDOS of states of (a1)-(f1) square and (g1)-(l1) hexagonal
nanodisks. Direction of the magnetic field is indicated by arrows.
Corresponding conductance of (a2)-(f2) square and (g2)-(l2) hexagonal
nanodisks are shown in the lower panels. We have calculated the conductance
between the A and B leads. The vertical axis is the bias energy, while the
vertical axis is the conductance in the unit of $e^2/h$. 
We have set $t=1;\protect\lambda =1, m=1, B=0.5$, and $t_{\ell}=t$. 
The sample size is $L=8$ for the square nanodisk and $L=6$ for the hexagonal nanodisk.}
\label{FigDevice}
\end{figure*}

\textit{Hamiltonians:} We start with the two-dimensional Hamiltonian\cite{EzawaNJP,MagHOTI} 
consisting of the hopping term $H_{t}$, the spin-orbit
interaction term $H_{\text{SO}}$ and the Zeeman term $H_{Z}$,
\begin{equation}
H_{\text{2D}}=H_{t}\tau _{z}+H_{\text{SO}}\tau _{x}+H_{Z},  \label{Hamil2D}
\end{equation}
with 
\begin{align}
H_{t}& =\sum_{n=1}^{N}m-t\sum \cos \left( \mathbf{d}_{n}\cdot \mathbf{k}\right) , \\
H_{\text{SO}}& =\lambda \sum_{n=1}^{N}C_{N}^{n}\sigma _{x}C_{N}^{-n}\sin
\left( \mathbf{d}_{n}\cdot \mathbf{k}\right) , \\
H_{Z}& =B\sigma _{x}\cos \theta +B\sigma _{y}\sin \theta
\end{align}
in the momentum space. We consider the square and triangular lattices with $N=4$ and $6$, respectively. 
Here, $m$, $t$, $\lambda $ are real parameters, $\mathbf{k}=\left( k_{x},k_{y}\right) $, 
and $\mathbf{d}_{n}=|\mathbf{d}_{n}|[\cos (2\pi n/N),\sin (2\pi n/N)]$; 
$\mathbf{\sigma }=(\sigma_{x},\sigma _{y},\sigma _{z})$ 
and $\mathbf{\tau }=(\tau _{x},\tau _{y},\tau_{z})$ represent 
the Pauli matrices for the spin and the pseudospin corresponding to the orbital degrees of freedom, 
respectively; $C_{N}=\tau_{0}\exp \left[ -i\pi \sigma _{z}/N\right] $ is the generator of the $\pi /N$ rotation. 
It reproduces the Bernevig-Huges-Zhang model\cite{BHZ} for $N=4$.
The Zeeman term may be introduced by applying inplane magnetic field or inplane magnetization.
Without the Zeeman term the system describes a time-reversal invariant topological insulator (TI).

\textit{Nanodisk:} We analyze a nanodisk geometry.
It is straightforward to calculate the local density of states (LDOS) by
changing the angle $\theta $ of the magnetic field as in Fig.\ref{FigDevice}(a1)--(f1) for a square nanodisk 
and as in Fig.\ref{FigDevice}(g1)--(l1) for a hexagonal nanodisk. 
It is observed that the LDOS is concentrated along the
edges. Hence, it is expected that the current follows mainly along the
edges. Indeed, we are able to calculate the conductance employing the method
we describe later: See Eq.(\ref{G}). We show the conductance between the two
leads attached to the corners A and B of a square nanodisk [Fig.\ref{FigDevice}(a1)] in Fig.\ref{FigDevice}(a2)--(f2), 
and that of a hexagonal nanodisk [Fig.\ref{FigDevice}(g1)] in Fig.\ref{FigDevice}(g2)--(l2).

First, we investigate a square nanodisk.
In the absence of the magnetic field, 
the topological edge states emerge for all sample edges as shown in Fig.\ref{FigDevice}(a1).
Once it is introduced into the $x$ direction, the edge states along the $y$ direction are gaped, 
while those along the $x$ direction remain gapless [Fig.\ref{FigDevice}(b1)]. 
When it is rotated away from the $x$ direction, the LDOS of the edge states is
distorted so that the maximum value takes at the corner of the sample.
The two corner states emerge when the magnetic field is along the $x+y$ direction [Fig.\ref{FigDevice}(d1)]. 
Finally, the edge states along the $x$ direction are gaped and those along the $y$ direction become gapless,
when the magnetic field becomes parallel to the $y$ direction.

\begin{figure}[t]
\centerline{\includegraphics[width=0.49\textwidth]{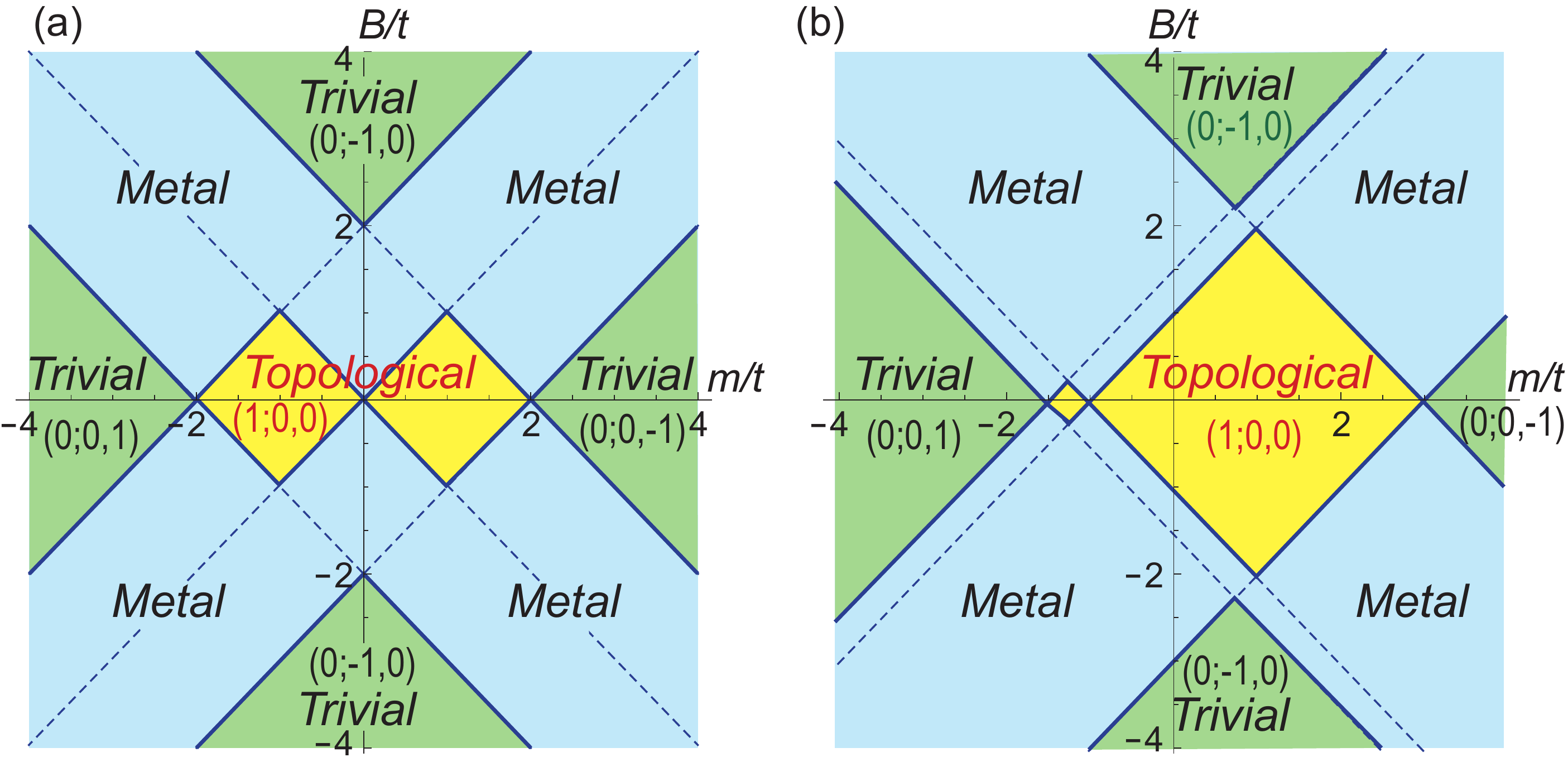}}
\caption{Topological phase diagrams for the (a) square and (b) hexagonal nanodisks. 
Dotted lines are determined by the zero-energy conditions at the high symmetry points.
Heavy curves are the phase boundaries.
The set ($\protect\nu_{\theta} $; $W_{\theta}$, $\protect\zeta $) of the
topological number $\protect\nu_{\theta} $, the mirror symmetry indicator $W_{\theta}$ and the
inversion symmetry indicator $\protect\zeta $ is well defined (a)  for the square nanodisk 
when $\theta=p\pi/4$ with integer $p$ satisfying $0\le p \le 7$ 
and (b) for the hexagonal nanodisk 
when $\theta=p\pi/6$ with integer $p$ satisfying $0\le p \le 11$.}
\label{FigPhase}
\end{figure}

These features are understood based on the low-energy effective Hamiltonians
of (\ref{Hamil2D}). They read
\begin{align}
H_{x} =&(k_{x}+B_{x})\sigma _{x}+B_{y}\sigma _{y}, \\
H_{y} =&(k_{y}+B_{y})\sigma _{y}+B_{x}\sigma _{x},
\end{align}
for the helical edges along the $x$ axis and the $y$ axis, respectively,
with $B_{x}=B\cos \theta $ and $B_{y}=B\sin \theta $. 
The energies are given by $E_{x}=\sqrt{(k_{x}+B_{x})^{2}+B_{y}^{2}}$ and $E_{y}=\sqrt{(k_{y}+B_{y})^{2}+B_{x}^{2}}$. 
These Hamiltonians imply the spin-momentum locking such that the spin polarization of the helical edge is locked parallel to  the edge for $B_x=0$ or $B_y=0$.

Indeed,  when $B_{x}\not=0$ and $B_{y}= 0$, 
$B_{x}$ opens the gap along the $y$ axis, while it induces only the shift of the helical edges along the $x$ axis. 
Consequently, there are topological edge channels along the $x$ axis. 
Each channel carries the unit conductance $e^{2}/h$ as in Fig.\ref{FigDevice}(b2). 
As we shall soon see, the system is a TCI protected by the mirror symmetry with respect to the $x$ direction and the inversion symmetry.

When $B_{x}\neq 0$ and $B_{y}\neq 0$, 
there are no topological edge states, implying the decrease of the conductance. 
The gap becomes largest both for the $x$ and $y$ directions when $B_{x}=$ $B_{y}\neq 0$, 
where the LDOS is concentrated at the two corners as in Fig.\ref{FigDevice}(d1).
We shall soon show that the system is a SOTI protected by the mirror symmetry with respect to the $x+y$ direction and the inversion symmetry.

Finally,  when $B_{x}=0$ and $B_{y}\neq 0$, the topological edge channels emerge along the $y$ direction [Fig.\ref{FigDevice}(f1)],
where the system is a TCI protected by the mirror symmetry with respect to the $y$ direction and the inversion symmetry.

We may similarly investigate the boundary states of the hexagonal nanodisks.
The results are shown in Fig.\ref{FigDevice}(g1)--(l2). 
Zero-energy channels appear along the $x$ axis when $\theta =0$ [Fig.\ref{FigDevice}(h1)]. 
All edges are gaped for $0<\theta <\pi/3 $. 
The SOTI is realized for $\theta =\pi /6$, where two topological corner states emerge [Fig.\ref{FigDevice}(j1).
Finally, topological edge channels appear when $\theta =\pi/3$ [Fig.\ref{FigDevice}(l1)]. 

\textit{Topological phase diagram: } We construct the phase diagram in the ($m/t,B/t$) plane. 
We first study the square nanodisk. The high symmetry points are 
$\Gamma =(0,0)$, $X=(\pi ,0)$, $Y=(0,\pi)$ and $M=\left( \pi ,\pi \right) $, 
where the energies are analytically given by
\begin{align}
E_{\Gamma }(0,0)=&2t+m\pm B,\quad -2t-m\pm B, \\
E_{M}(\pi ,\pi)=&2t-m\pm B,\quad -2t+m\pm B,\\
E_X(\pi ,0=&E_Y\left( 0,\pi \right) =m\pm B,-m\pm B,
\end{align}
which are independent of $\lambda$ and $\theta $. 
Insulator and metallic regions are determined by diagonalizing the Hamiltonian.
It turns out that the phase boundaries are well reproduced by the zero-energy conditions $E_{\Gamma }=E_{M}=E_{X}=0$
at the high symmetry points. 
Insulator regions emerge when Dirac cones are fixed at these points,
while metallic regions emerge when Dirac cones move along the line $M$-$X$-$M$. 
We will soon assign the topological number to each insulator phase: See Fig.\ref{FigPhase}(a).
It is known\cite{Schin} that, when $B=0$, the system is a topological insulator for $\left\vert m/t\right\vert <2$ 
and a trivial insulator for $\left\vert m/t\right\vert <2$,
as agrees with the present result.

We may similarly construct the phase diagram for the hexagonal nanodisk. 
The high symmetry points are $\Gamma=( 0,0) $, $M=(0,2\pi /\sqrt{3}) $  and $K=(2\pi /3,0)$, 
where the energies are analytically given by
\begin{align}
E_{\Gamma }(0,0) =&3t-m\pm B,\quad -3t+m\pm B, \\
E_{M}(0,2\pi /\sqrt{3}) =&t+m\pm B,\quad -t-m\pm B, \\
E_{K}(2\pi /3,0) =&\frac{3}{2}t+m\pm B,\quad -\frac{3}{2}t-m\pm B.
\end{align}
The phase boundaries are well described by the zero-energy conditions $E_{\Gamma } =E_{M}=E_{K}=0$ as in Fig.\ref{FigPhase}(b).

\begin{figure}[t]
\centerline{\includegraphics[width=0.49\textwidth]{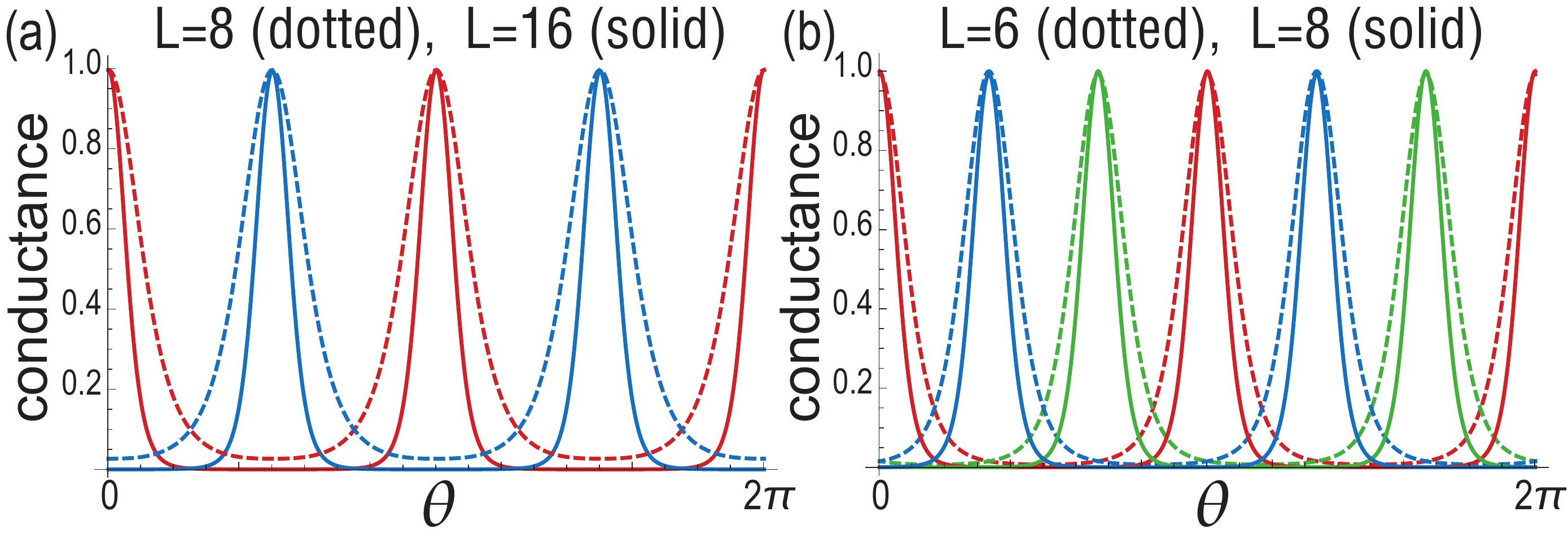}}
\caption{Angle dependence of the conductance for the (a) square and (b)
hexagonal nanodisks. (a) Red (blue) curves represent the conductance between
A and B (A and D) in Fig.\protect\ref{FigDevice}(a1). (a) Red (blue) [green]
curves represent the conductance between A and B (A and F) [B and C] in Fig.\protect\ref{FigDevice}(g1). 
The horizontal axis is the angle of the magnetic
field, while the vertical axis is the conductance. }
\label{FigRotat}
\end{figure}

\begin{figure*}[t]
\centerline{\includegraphics[width=0.85\textwidth]{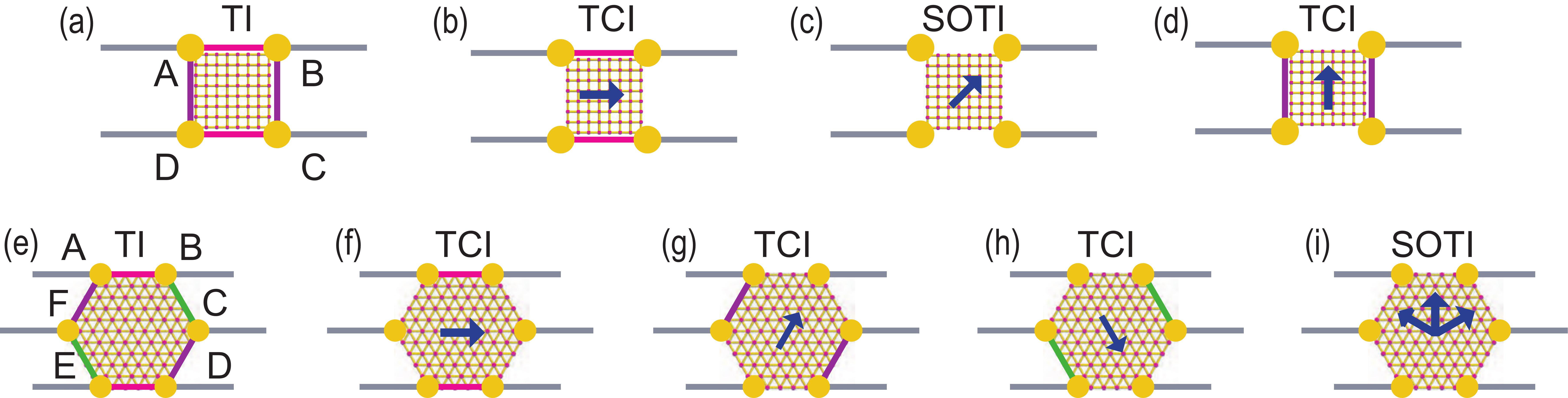}}
\caption{Schematic figures of topological circuit changing switches, where
topological edge states connect two leads. The connectivity can be
controlled by changing the magnetic field angle. 
The topological circuits are switched on (off) for the TCI (SOTI) phase.
We can make a two step switch for the square nanodisk and a three step switch for the hexagonal nanodisk. }
\label{FigSwitch}
\end{figure*}

\textit{Mirror symmetry indicator: }We explore the symmetry indicators to characterize the bulk topology of the square nanodisk. 
We search for the mirror symmetry operator $M_{\theta}$ by changing the direction $\theta$ of the magnetic field.
The mirror symmetries are present at $\theta=p\pi/4$ with $p$ being integers such that $0\le p \le 7$. 
The mirror eigenvalue is given by solving $M_{\theta}|\psi_n (k)\rangle =\chi_n (k)|\psi_n (k)\rangle $ with the band index $n$, and
it is quantized as $\chi_n (k)=\pm i$ along the mirror-symmetric line because $M^2_{\theta}=-1$.
The winding number along the mirror symmetric line is defined as 
\begin{equation}
W_{\theta}=\frac{i}{4\pi }\sum_n\int_{-\pi }^{\pi }\chi_n (k)dk,  \label{chi}
\end{equation}
where $n$ runs over the occupied bands.
It is quantized since $\chi_n (k)$ is a constant.
This is the mirror symmetry indicator, which reads in each phase as in Fig.\ref{FigPhase}(a).

First, for $\theta=0$, there is a mirror symmetry $M_{0}=i\sigma_{x}$ with respect to the $x$ direction. 
Topological edge states emerge parallel to the $x$ axis as in Fig.\ref{FigDevice}(b1), which indicates that the system is a TCI.
TCIs are realized also at $\theta=\pm \pi/2, \pi$.

Second, for $\theta=\pi/4$, there is a mirror symmetry $M_{\pi/4}=i(\sigma_{x}+\sigma _{y})/\sqrt{2}$ with respect to the $x+y$ direction. 
Topological boundary states emerge at the two corners as in Fig.\ref{FigDevice}(d1), which indicates that the system is a SOTI.
SOTIs are realized also at $\theta=-\pi/4,\pm3\pi/4$.

Similar analysis is carried out for the hexagonal nanodisk,
where the mirror symmetry indicator is given by the same formula as (\ref{chi}) at
$\theta=p\pi/6$ with integer $p$ satisfying $0\le p \le 11$.
The results are shown in the phase diagram Fig.\ref{FigPhase}(b).

\textit{Inversion symmetry indicator:} The Hamiltonian (\ref{Hamil2D}) for
the square lattice has the inversion symmetry $I=\tau _{z}$ irrespective to
the magnetic field direction. The inversion eigenvalue $\xi $ of the
symmetry operator $I$, $I|\psi_n (k)\rangle =\xi_n(k) |\psi_n (k)\rangle $, is quantized\cite{Schin,CFang,k4,Kha} 
to be $\xi_n(k)=\pm 1$ at the inversion symmetric points because $I^{2}=1$. 
We define the $\mathbb{Z}$ index $\zeta $ protected by the inversion symmetry $I$ by the formula 
\begin{equation}
\zeta =\frac{1}{2}\sum_n\xi _{n}(k=\Gamma )+\frac{1}{2}\sum_n\xi _{n}(k=M),  \label{iota}
\end{equation}
where $n$ runs over the occupied bands at the points $\Gamma $ and $M$. 
We show the inversion symmetry indicator in each phase of the phase diagram in Fig.\ref{FigPhase}(a).

Similar analysis is carried out for the hexagonal nanodisk,
where the inversion symmetry indicator is given by the same formula as (\ref{iota}).
The results are shown in the phase diagram Fig.\ref{FigPhase}(b).

\textit{Bulk topological number:} Both in the square and hexagonal nanodisks 
the bulk topological number is defined with the use of the mirror and inversion symmetry indicators as
\begin{equation}
\nu_{\theta} =\text{mod}_{2}(\zeta -1)\text{mod}_{2}(W_{\theta}-1).  \label{TopNum}
\end{equation}
This is the $\mathbb{Z}_{2}$ number, which takes $\nu_{\theta}  =1$ for the
topological phase and $\nu_{\theta} =0$ for the trivial phase. We have shown them in
the phase diagrams Fig.\ref{FigPhase}(a) and (b).

\textit{Conductance:} The conductance between two leads is calculated as follows. 
The natural framework for transport calculations in nanoscopic devices is the
Landauer formalism\cite{Datta,Rojas,Nikolic,Li}. In terms of single-particle
Green's functions, the low-bias conductance $\sigma (E)$ at the energy $E$ is given by\cite{Datta} 
\begin{equation}
\sigma (E)=(e^{2}/h)\text{Tr}[\Gamma _{\text{L}}(E)G_{\text{D}}^{\dag
}(E)\Gamma _{\text{R}}(E)G_{\text{D}}(E)],  \label{G}
\end{equation}
where $\Gamma _{\text{R(L)}}(E)=i[\Sigma _{\text{R(L)}}(E)-\Sigma _{\text{R(L)}}^{\dag }(E)]$ 
with the self-energies $\Sigma _{\text{L}}(E)$ and $\Sigma _{\text{R}}(E)$, and
\begin{equation}
G_{\text{D}}(E)=[E-H_{\text{D}}-\Sigma _{\text{L}}(E)-\Sigma _{\text{R}}(E)]^{-1},  \label{StepA}
\end{equation}
with the Hamiltonian $H_{\text{D}}$ for the device region. The self energy
of single-atomic semi-infinite chain is analytically obtained\cite{Datta} as 
\begin{equation}
\Sigma _{\text{L}}(E)=\Sigma _{\text{R}}(E)=E-i\sqrt{\left\vert t_{\ell
}^{2}-E^{2}\right\vert },
\end{equation}
where $t_{\ell }$ is the transfer integral of the single-atomic
semi-infinite chain.

We show the conductance as a function of the energy $E$ in Fig.\ref{FigDevice} 
both for the square and hexagonal nanodisks. We have
calculated the conductance between the leads A and B. 
Its behaviors read as follows. It is almost constant when the energy $E$ is within the bulk gap $\Delta $. 
On the other hand, the conductance fluctuates rapidly when the energy $E$ is in the bulk band. 
This is because the leads are too
narrow to convey the conductance of the bulk.

\textit{Topological devices:} We consider topological devices, where four
(six) leads are attached to the corners of the square (hexagonal) nanodisk: See Fig.\ref{FigIllust}. 
The leads are single atomic chains with semi-infinite length. 
Experimentally, the leads can be replaced by multi-terminal STM  (Scanning Tunneling Microscope) chips.

The conductance between two leads is shown as a function of $\theta $ in Fig.\ref{FigRotat}. 
As a pair of two leads we have taken (A,B) and (A,D) for the
square nanodisk [Fig.\ref{FigSwitch}(a)]; (A,B), (A,F) and (B,C) for
the hexagonal nanodisk [Fig.\ref{FigSwitch}(e)]. The conductance
rapidly changes as $\theta $ increases. Hence, we can use it as a
topological circuit changing switch. In another view point, we can detect
the direction of magnetization of a sample surface by measuring the
conductance. It is a topological magnetic sensor. The angle dependence of
the conductance becomes sharper for larger nanodisks. A remarkable property
is that it is enough to use a nanodisk as small as only 10nm for a clear
switching of the conductance.

The author is very much grateful to S. Hasegawa, T. Hirahara and N. Nagaosa
for helpful discussions on the subject. This work is supported by the
Grants-in-Aid for Scientific Research from MEXT KAKENHI (Grants No.
JP17K05490, No. JP15H05854 and No. JP18H03676). This work is also supported
by CREST, JST (JPMJCR16F1).


\begin{thebibliography}{99}


\bibitem{FuTCI} L. Fu, Phys. Rev. Lett. \textbf{106}, 106802 (2011).

\bibitem{AndoReview} Y Ando, L Fu, Annu. Rev. Condens. Matter Phys. 6 (1), 361 (2015)

\bibitem{Fan} F. Zhang, C.L. Kane and E.J. Mele, Phys. Rev. Lett. \textbf{110}, 046404 (2013).

\bibitem{Science} W. A. Benalcazar, B. A. Bernevig, and T. L. Hughes, 
\textbf{357}, 61 (2017).

\bibitem{APS} F. Schindler, A. Cook, M. G. Vergniory, and T. Neupert, in APS
March Meeting (2017).

\bibitem{Peng} Y. Peng, Y. Bao, and F. von Oppen, Phys. Rev. B \textbf{95},
235143 (2017).

\bibitem{Lang} J. Langbehn, Y. Peng, L. Trifunovic, F. von Oppen, and P. W.
Brouwer, Phys. Rev. Lett. \textbf{119}, 246401 (2017).

\bibitem{Song} Z. Song, Z. Fang, and C. Fang, Phys. Rev. Lett. \textbf{119},
246402 (2017).

\bibitem{Bena} W. A. Benalcazar, B. A. Bernevig, and T. L. Hughes, Phys.
Rev. B \textbf{96}, 245115 (2017).

\bibitem{Schin} F. Schindler, A. M. Cook, M. G. Vergniory, Z. Wang, S. S. P.
Parkin, B. A. Bernevig, and T. Neupert, cond-mat/arXiv:1708.03636 (2017).

\bibitem{FuRot} C. Fang, L. Fu, arXiv:1709.01929.

\bibitem{EzawaKagome} M. Ezawa, Phys. Rev. Lett. \textbf{120}, 026801 (2018).

\bibitem{EzawaPhos} M. Ezawa, arXiv:1801.00437.

\bibitem{Gei} M. Geier, L. Trifunovic, M. Hoskam, and P. W. Brouwer,
arXiv:1801.10053.

\bibitem{MagHOTI} M. Ezawa, Phys. Rev. B \textbf{97}, 155305 (2018).

\bibitem{Kha} E. Khalaf, Phys. Rev. B \textbf{97}, 205136 (2018).

\bibitem{HexaHOTI} M. Ezawa, arXiv:1803.02995; Phys. Rev. B , (R) (2018), accepted for publication.

\bibitem{Bis} F. Schindler, Z. Wang, M. G. Vergniory, A. M. Cook, A. Murani,
S. Sengupta, A. Y. Kasumov, R. Deblock, S. Jeon, I. Drozdov, H. Bouchiat, S.
Gueron, A. Yazdani, B. A. Bernevig, and T. Neupert, arXiv:1802.02585.

\bibitem{EzawaNJP} M. Ezawa, New J. Phys. \textbf{16}, 065015 (2014).

\bibitem{BHZ} B. A. Bernevig, T. L. Hughes, and S.-C. Zhang, Science \textbf{314},
1757 (2006).

\bibitem{CFang} C. Fang, Z. Song, and T. Zhang, arXiv:1711.11050.

\bibitem{k4} E. Khalaf, H. C. Po, A. Vishwanath and H. Watanabe,
arXiv:1711.11589.

\bibitem{Datta} S. Datta, \textit{Electronic Transport in Mesoscopic Systems}
(Cambridge University Press, Cambridge, England, 1995): \textit{Quantum
transport: atom to transistor} (Cambridge University Press, England, 2005).

\bibitem{Rojas} F. Mu\~{n}oz-Rojas, D. Jacob, J. Fern\'{a}ndez-Rossier, and
J. J. Palacios, Phys. Rev. B \textbf{74}, 195417 (2006).

\bibitem{Nikolic} L. P. Z\^{a}rbo and B. K. Nikoli\'{c}, , EPL, 80 47001
(2007): D. A. Areshkin and B. K. Nikoli\'{c}, Phys. Rev. B \textbf{79},
205430 (2009).

\bibitem{Li} T. C. Li and S.-P. Lu, \textit{Quantum conductance of graphene
nanoribbons with edge defects},\ Phys. Rev. B \textbf{77}, 085408 (2008).
\end{thebibliography}
\end{document}